\documentclass[12pt]{iopart}

\usepackage{amsbsy}
\usepackage{latexsym}
\usepackage{graphics}
\usepackage[squaren,thickspace,thickqspace]{SIunits}
\usepackage{multirow}
\usepackage{textcomp}
\usepackage{amsfonts}
\usepackage[numbers]{natbib}
\usepackage{ifthen}

\newcommand{\be}{\begin{eqnarray}}
\newcommand{\ee}{\end{eqnarray}}

\newcommand{\Beplus}{\ensuremath{{^9}{\rm Be}^{+} \,}}

\newcommand{\D}{\mathrm{d}}
\newcommand{\I}{\mathrm{i}}
\newcommand{\Exp}[1]{\mathrm{e}^{#1}}
\newcommand{\der}[3][1]{\ifthenelse{#1=1}{\frac{\D#2}{\D#3}}
	{\frac{\D^#1#2}{\D#3^#1}}}%
\newcommand{\pder}[3][1]{\ifthenelse{#1=1}
	{\frac{\partial #2}{\partial #3}}{\frac{\partial ^#1#2}{\partial #3^#1}}}%
\renewcommand{\vec}[1]{{\boldsymbol{#1}}}
	\renewcommand{\phi}{\varphi}
\let\FPi\Pi	\renewcommand{\Pi}{\mathit{\FPi}}

\begin{document}

\title{Optimal electrode geometries for 2-dimensional ion arrays with bi-layer ion traps}

\author{F N Krauth, J~Alonso and J~P~Home}

\address{Institute for Quantum Electronics, ETH Z\"urich, Otto-Stern-Weg 1, 8093 Z\"urich, Switzerland}

\ead{jhome@phys.ethz.ch}

\begin{abstract}
We investigate electrode geometries required to produce periodic 2-dimensional ion-trap arrays with the ions placed between two planes of electrodes. We present a generalization of previous methods for traps containing a single electrode plane to this new geometry, and show that for a given ion-electrode distance and applied voltages, the inter-ion distance can be reduced by a factor of up to 3 relative to single-plane traps. This represents an increase by a factor of 9 in the trap density and a factor of 27 in the exchange coupling between the oscillatory motion of neighboring ions. The resulting traps are also considerably deeper for bi-layer structures than for single-plane traps. These results could offer a useful path towards 2-dimensional ion arrays for quantum simulation. We also discuss issues with the fabrication of such traps.
\end{abstract}

\pacs{pacs}
\maketitle

\section{Introduction}
\label{intro}
Lattice models play an important role in the study of many-body physics \cite{BkChaikin}. The recent improvements in the quantum state control of trapped atomic ions make these a leading system for precisely controlled realization of quantum simulation of lattice models \cite{04Porras2,06Porras,08Friedenauer,12Schneider,12Blatt,13Islam,13Richerme,14Sterling}. These could be realized by building 1 or 2-dimensional arrays of microtraps, where a single ion is trapped at each site. The long-range Coulomb interaction between the ions couples the motion of the ions at different traps, providing a boson hopping term between different sites. With the introduction of non-linearities, this can be used to investigate interacting boson models. Furthermore, the Coulomb coupling is a critical element for mediating effective spin-spin interactions between the internal states of the ions by introducing additional optical or microwave fields. In addition, the combination of motional and spin degrees of freedom provides opportunities for investigating generalized spin-boson models such as the Jaynes-Cummings-Hubbard model \cite{06Hartmann,06Greentree,07Angelakis,09Ivanov,13Toyoda}.

A promising direction for producing configurable trapped-ion lattices is to use periodic arrays of electrode structures placed on a single surface above which the ions are trapped \cite{05Chiaverini2,09Schmied}. The lattice sites are defined by the geometry of the electrodes and the applied potentials. One significant challenge in realizing such a system is to engineer exchange couplings for phonon hopping between neighboring sites which are large compared to the heating rates of the ions' motion. In all the cases we consider below, the exchange coupling for two identical ions of mass $M$ and charge $e$ oscillating along the same direction with a secular angular frequency $\omega$ in two traps separated by a distance $d$ is given to lowest order by
\be\label{eq:Jexgen}
\Omega_{\rm ex} \approx \frac{e^2}{4 \pi \epsilon_0}\frac{1}{2 M \omega}\left.\pder[2]{G(h,\rho)}{\rho}\right|_{\rho=d}.
\ee
Here $\rho^2=x^2+y^2$ and $x,y,z$ are the spatial coordinates. $G(h,\rho)$ is the Green's function for the system comprised by ions and electrodes given the ions are located on the plane $z=h$. Far from any conducting surface, the free-space Green's function is $G_\text{fs}(h,d)=1/d$ and we retrieve
\be\label{eq:Jex}
\Omega_{\rm ex,fs} \approx \frac{e^2}{4 \pi \epsilon_0 M \omega d^3},
\ee
which has the $d^{-3}$ scaling associated with dipole-dipole interactions \cite{11Brown,11Harlander,14Wilson}. This should be compared with heating caused by fluctuations in the electric field at the ion position, which scale rapidly with the ion-electrode distance. Experiments with small-scale ion traps exhibit heating rates well above that expected due to Johnson noise in the supply circuitry \cite{00Turchette}. Recent studies have shown reductions of the heating rate by factors of $\sim100$ by ablation of the surface using argon-ion guns \cite{12Hite,13Daniilidis}, and reductions by similar amounts by cryogenic cooling of the trap \cite{06Deslauriers,08Labaziewicz}. Assuming a model involving microscopic fluctuating voltages (or dipoles) on the electrode surface with a correlation length much smaller than the ion-electrode distance $h$, the predicted scaling is $\dot{\bar{n}}\propto M^{-1}\omega^{-1}h^{-4}$ \cite{00Turchette,11Danilidis,11Safavi}, where $\bar{n}$ is the mean motional quantum number and the dot represents the time derivative. This behavior is consistent with experimental observations \cite{06Deslauriers,14Chiaverini}. It can be seen from the scalings above that for this model $\Omega_{\rm ex}/\dot{\bar{n}}\propto h^4/d^3$, meaning it is desirable to keep the ions as far as possible from the electrode surface while having them packed as densely as possible for strong ion-ion coupling. Therefore, optimized electrode geometries which minimize $d$ for a given $h$ can offer important advantages. An additional reason to aim for large $h/d$ is ease of access for laser beams, reducing problems with light scattering as well as electric charging of dielectric materials or contaminants on the electrode surfaces.

Optimal electrode geometries for traps with a single plane of electrodes were given by Schmied \emph{et al.} \cite{09Schmied}, as well as for an array of electrodes with a single, grounded cover plane over the whole trap \cite{11Schmied}. In this paper, we extend these methods to an alternative electrode geometry in which ions are trapped between two planes of periodically structured electrodes. In section \ref{sec:simulation} we present the mathematical formalism employed to optimize such bi-layer trap configurations. In section \ref{sec:opt_struc} we show that this allows for a significant reduction in the inter-ion distance for a given ion-electrode distance. The traps resulting from bi-layer configurations are also considerably deeper than in the case of a single plane of electrodes. In this section we also give some examples of electrode geometries and quantify the improvement in trap density with respect to the single-plane case. An experimental challenge would be to fabricate such structures, in particular realizing sufficient relative alignment of the two planes. This and other considerations are discussed in sections \ref{sec:fab} and \ref{sec:disc}.

\section{Simulation of multiple electrode planes}
\label{sec:simulation}

\begin{figure}
\centering
\resizebox{0.5\textwidth}{!}{
  \includegraphics{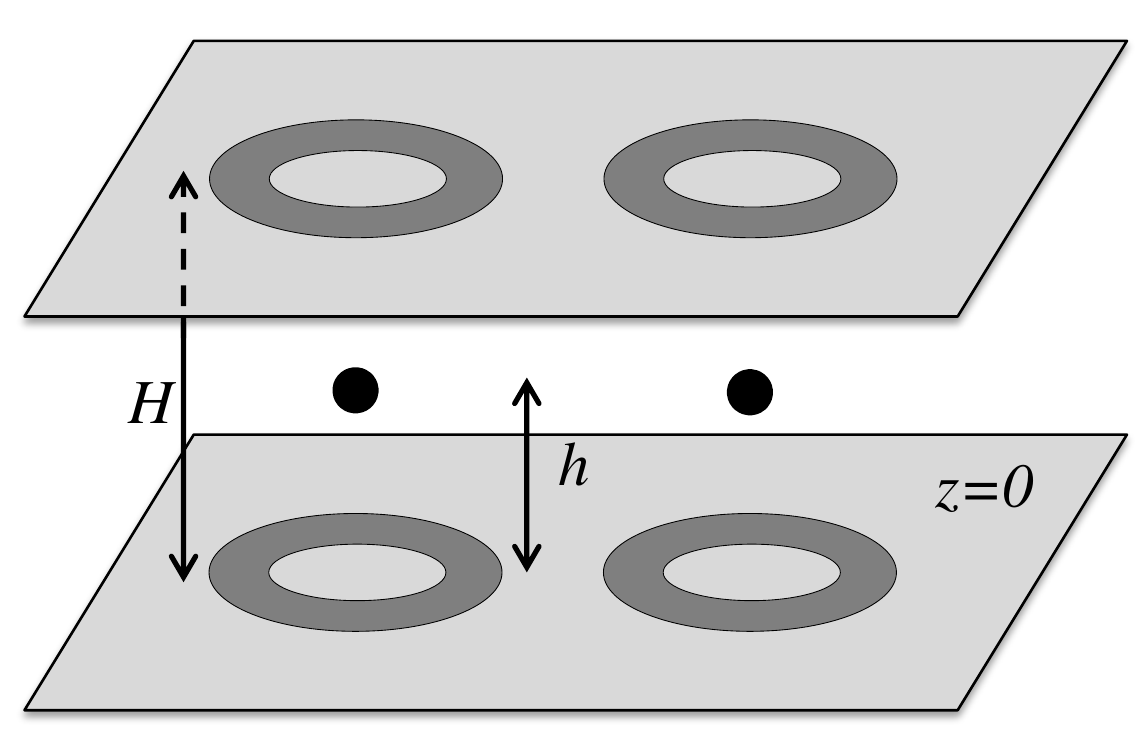}
}
\caption{Depiction of bi-layer trap. It consists of two infinite grounded planes (light gray) separated by a distance $H$. On them, 2-dimensional patterns define electrodes (dark gray) onto which radio-frequency voltages can be applied to create trapping locations (black dots) at a height $z=h$.}
\label{fig:bilayer_trap}
\end{figure}

We want to characterize infinite periodic arrays of traps created in the plane $z=h$ by applying radio-frequency potentials to electrodes patterned on two parallel planes at $z=0$ and $z=H$ (see figure \ref{fig:bilayer_trap}). The principal observation which we make is that such bi-layer traps can be analyzed in a similar manner to the traps produced by an infinite periodic plane of electrodes at $z=0$ with a parallel grounded plane at $z=H$, using the principle of superposition. In reference \cite{11Schmied} it is shown that, in the case of a single electrode plane at $z'=0$, the potential at any point $\vec{r}=(x,y,z)$ is
\be
\Phi(\vec{r})=\displaystyle\sum_{k_x,k_y}\tilde{\Phi}(k_x,k_y)\tilde{G}^{(S)}(k_x,k_y,z)\Exp{\I (k_xx+k_yy)}.
\ee
Here, $k_i=2\pi n_i/L_i$ and $k=({\sum_i{k_i^2}})^{1/2}$, with $n_i\in \mathbb{Z}$ and $L_i$ the spatial period along the $i$-direction. $\tilde{\Phi}(k_x,k_y)$ is the Fourier transform of the surface potential on the electrode plane,
\be
\tilde{\Phi}(k_x,k_y)=\frac{1}{2\pi}\int\Phi(x,y,0)\Exp{-\I (k_xx+k_yy)}\D x\D y,
\ee
and $\tilde{G}^{(S)}(k_x,k_y,z)$ is the Fourier transform of a ``surface Green's function'', defined by $G^{(S)}(\vec{r},\vec{r'})\equiv\frac{1}{4\pi}\left.\pder{G(\vec{r},\vec{r'})}{z'}\right|_{z'=0}$, and can be written as
\be
\tilde{G}^{(S)}(k_x,k_y,z)=\frac{\sinh(kH-kz)}{\sinh(kH)}.
\ee

The total potential between the planes of a bi-layer trap can be expressed as the sum of two terms: $\Phi_\text{tot}(\vec{r})=\Phi_1(\vec{r})+\Phi_2(\vec{r})$. $\Phi_1$ is contributed by the surface potentials on the segmented array at $z=0$ with the grounded infinite plane a distance $H$ above it. $\Phi_2$ arises from the surface potentials on a segmented array at $z=H$ with an infinite grounded plane at $z=0$. To calculate $\Phi_2$ we use the equations above after applying the transformation $z\rightarrow H-z$.

From this starting point it becomes possible to solve for multi-layer trap geometries. The algorithm determines which of the individual segments on each of the planes should be grounded and which connected to the trap voltage drive. We will see below that this leads to identical patterns on both planes. However, this is not enforced by the code, which is an extended version of the \emph{SurfacePattern} package provided by Schmied \cite{SchmiedSoftware}.

\section{Optimized electrode structures}
\label{sec:opt_struc}

We consider an electrode geometry optimal if it maximizes the dimensionless curvature $\kappa=\frac{\sqrt{2}M\tilde{\omega}\omega_\text{rf}h^2}{eU_\text{rf}}$ of the potential minima it generates. $U_\text{rf}$ and $\omega_\text{rf}$ are the amplitude and angular frequency of the radio-frequency potentials applied to the electrodes respectively, and $\tilde{\omega}$ is the mean motional angular frequency (geometric mean over the three principal axes of the trap). We then analyze the resulting electrode geometry with regard to trap depth, which is important because deep traps are easier to load and can store the ions for longer times. To parametrize the trap depth $\Psi$ we use the dimensionless parameter $\eta=\frac{4M\omega_\text{rf}^2h^2\Psi}{e^2U_\text{rf}^2}=\frac{2\kappa^2\Psi}{M\tilde{\omega}^2h^2}$. Both $\kappa$ and $\eta$ depend solely on the electrode geometry (see \cite{09Schmied}).

\subsection{Optimal single traps}

For the case of a single trap and no cover plane ($H\rightarrow\infty$), the optimal geometry is a ring electrode of inner radius $r_1=0.678h$ and outer radius $r_2=3.381h$, giving $\kappa=0.298$ and $\eta=0.020$ \cite{09Schmied}. If we instead consider a single trap with electrodes on two opposing planes, the optimal geometry is a mirror-symmetric configuration with a disc of radius $r=0.800h$ (and $H=2h$). This already suggests that placing electrodes on a cover plane offers the possibility of increasing the lateral density of traps. Furthermore, the dimensionless parameters for such a ``double-disc trap'' are $\kappa=0.672$ and $\eta=0.080$, a factor of 2.3 and 4 higher than those of the ``single ring trap'' respectively. The reason for such an improvement in the trap depth is that the lowest energy saddle-point for the double-disc trap is no longer along the $z$-axis (as is typical in surface-electrode traps \cite{05Chiaverini2,11Amini}), but in-plane. It should be noted that this improvement will not be as large if compared to a surface-electrode trap with a grounded cover plane \cite{11Schmied}. In this case, the optimal ring trap has radii $r_1=0.81h$ and $r_2=17.2h$, $H=3.17h$ and the trap parameters are $\kappa=0.350$ and $\eta=0.076$. The cover-plane height is much lower than is usual in current experimental setups ($\sim\text{mm}$). This is due to a difference in function. Whereas the cover plane introduced here attempts to produce traps with higher curvature, the cover planes more typically used in experimental setups are for increasing the trap depth during ion loading \cite{09Clark}. It should be noted the latter procedure can displace the potential minimum from the pseudo-potential null. We note also that the large size of the optimal rings make dense packing difficult in this configuration.

\subsection{Periodic arrays of traps}

\begin{figure}
\centering
\resizebox{0.8\textwidth}{!}{
  \includegraphics{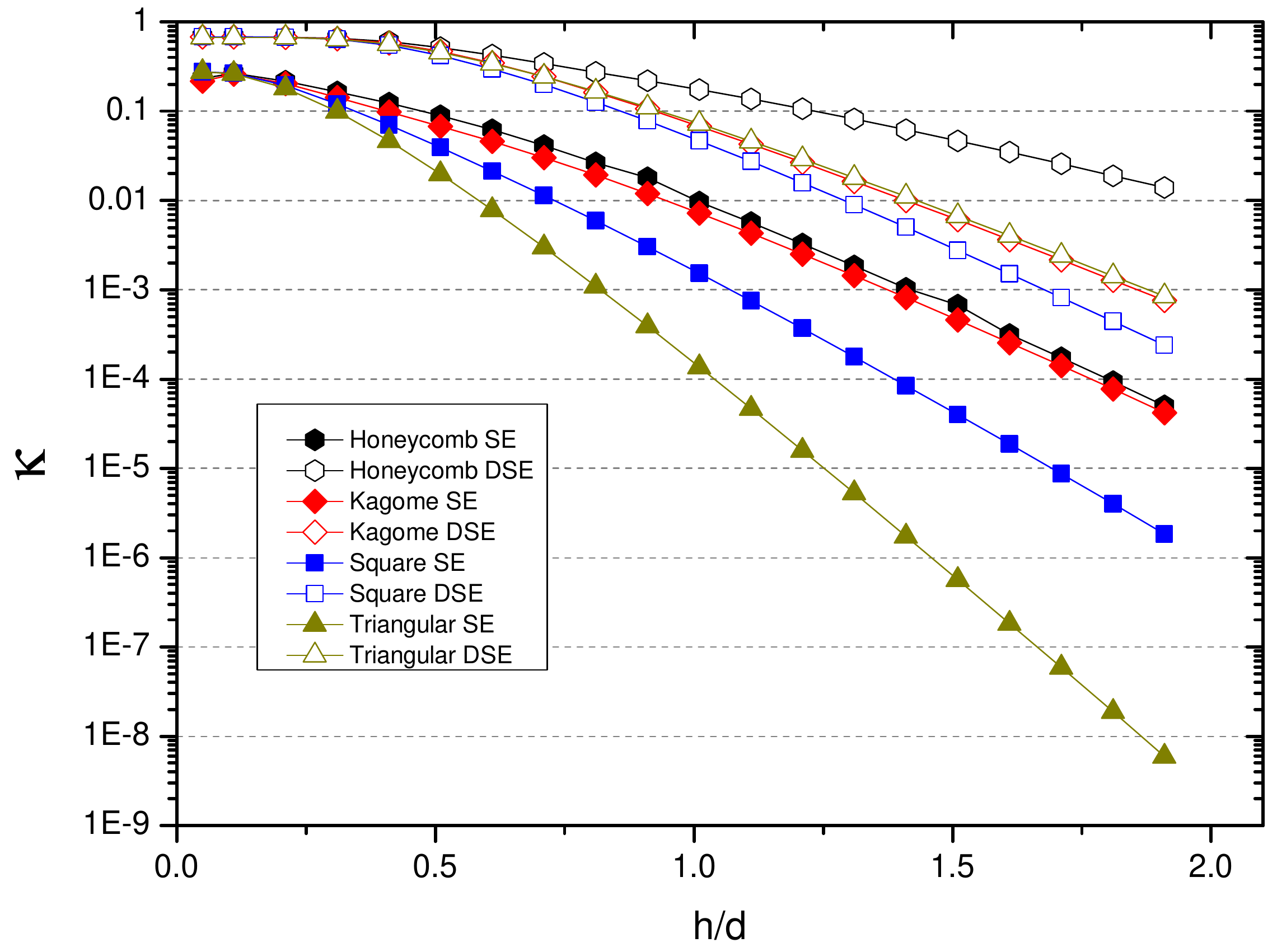}
}
\caption{Comparison of  the dimensionless curvature $\kappa$ for bi-layer (open symbols) and surface-electrode (filled symbols) configurations. The data for the Kagome and triangular lattices in the bi-layer configurations overlap in this plot.}
\label{fig:kappa}
\end{figure}

\begin{figure}
\centering
\resizebox{0.8\textwidth}{!}{
  \includegraphics{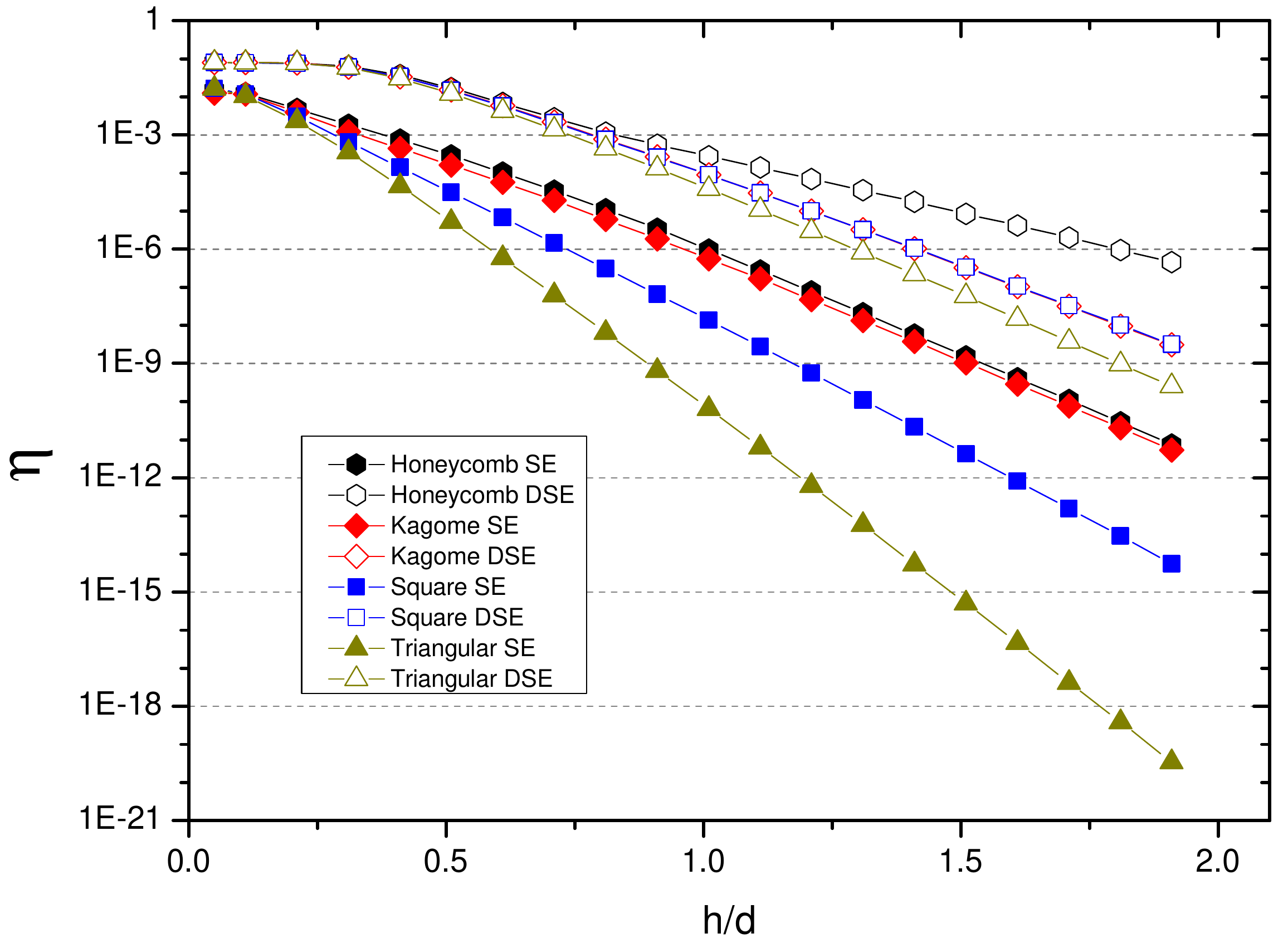}
}
\caption{Comparison of the dimensionless trap depth $\eta$ for bi-layer (open symbols) and surface-electrode (filled symbols) configurations. The data for the Kagome and square lattices in the bi-layer configurations overlap in this plot.}
\label{fig:eta}
\end{figure}

\begin{figure}
\centering
\resizebox{0.8\textwidth}{!}{
  \includegraphics{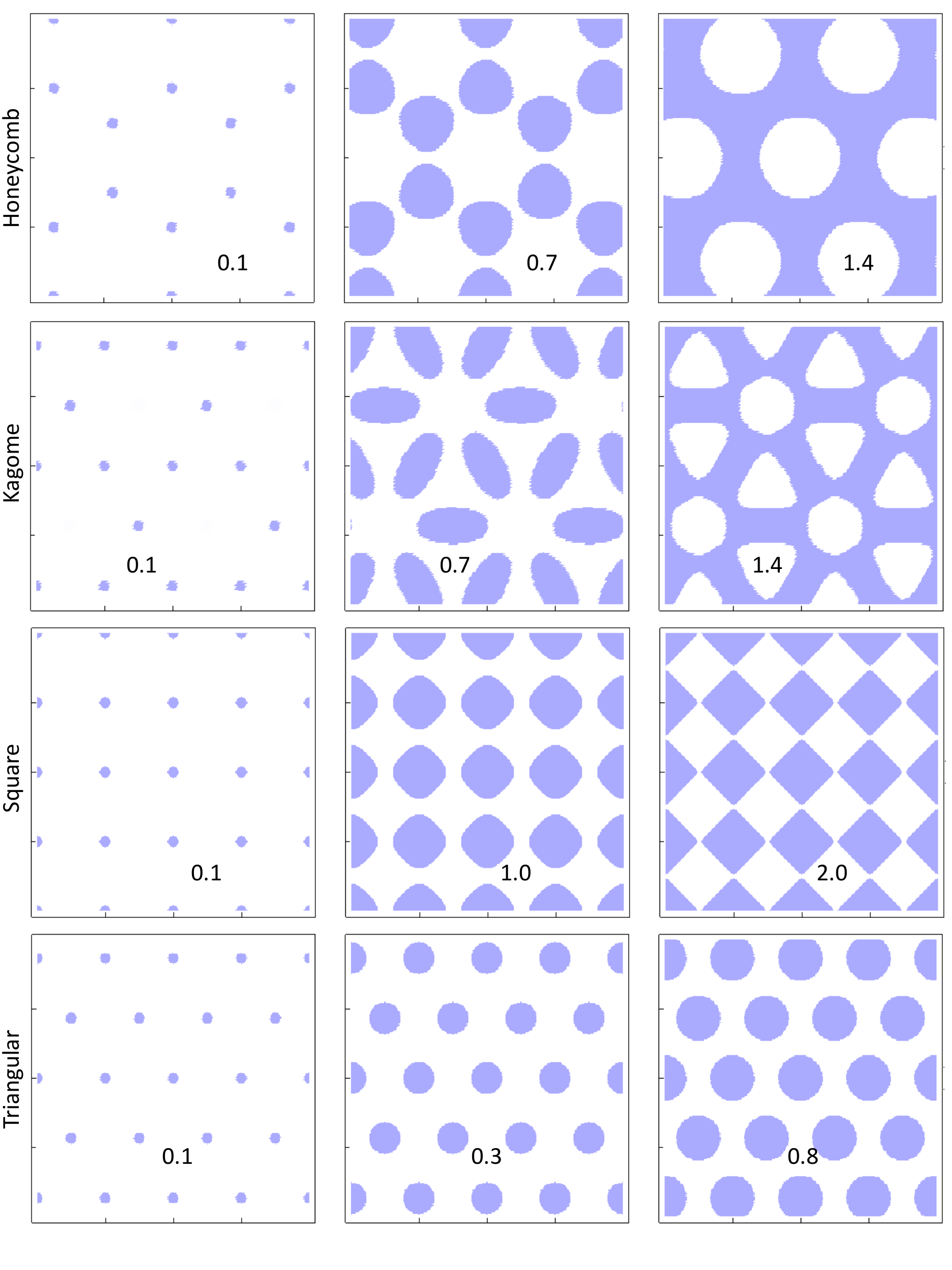}
}
\caption{Electrode patterns for the bottom surface of bi-layer traps at variable ratio $h/d$ (indicated in each plot) for different lattices. The radio-frequency voltage is applied on the shaded structures, while white ones remain grounded. Because the dimensionless curvature is maximal when the trap plane is at half the height of the cover plane, the trap is symmetric with respect to that mirror plane and the patterns are also symmetric. For small $h/d$ the electrode patches are discs under the trap location. As $h/d$ grows the electrodes grow until they start to interfere and patches change their shape (except for the triangular lattice). For each lattice the third pattern was chosen to be the maximally interfering pattern: increasing the height will no longer change the pattern (higher voltages will be required in order to keep the curvature constant).}
\label{fig:patterns}
\end{figure}

Figures \ref{fig:kappa} and \ref{fig:eta} show the results of the optimization for infinite periodic arrays of traps, both for surface-electrode and bi-layer traps. The lattices simulated are honeycomb, Kagome, square and triangular (see first column in figure \ref{fig:patterns}), and we have forced both principal in-plane axes of the traps to lie along the $x$ and $y$ directions which determine the lattice orientation. In the regime where the ion-ion distance $d$ is much larger than the ion-electrode distance $h$, the optimal electrode configuration is simply an array of the optimal single traps (ring traps for a single plane, double-disc traps for the bi-layer case). As $h/d$ increases, the geometries of neighboring traps begin to interfere until the point is reached where single patterns lack space for expansion and geometries remain unchanged. Figure \ref{fig:patterns} shows the resulting electrode patterns for the different lattices for a few chosen values of $h/d$ in the bi-layer configuration. The rightmost pattern is the one to which the lattice converges. In the specific case of the square lattice, spurious traps appear in the center of the squares and can be very similar to the main traps, effectively reducing the lattice constant. They only become identical in the limit $h/d\rightarrow\infty$, so our optimization code does not detect them, but they could be taken advantage of in practical realizations.

The lattices and conditions chosen allow for a direct comparison with the results from \cite{09Schmied}. For example, if we aim at a value of $\kappa=0.1$ we see that we are limited to $h/d$ values between 0.3 and 0.5 for the surface-electrode traps, while we can reach values from 0.85 to 1.25 with bi-layer traps. In this case, the improvement ranges from a factor slightly above 2 (square and Kagome lattices) to around 3 (triangular lattice). That means an increase by a factor of 4 to 9 in the trap density and a factor of 8 to 27 in the coupling between ions.

\begin{figure}
\centering
\resizebox{0.8\textwidth}{!}{
  \includegraphics{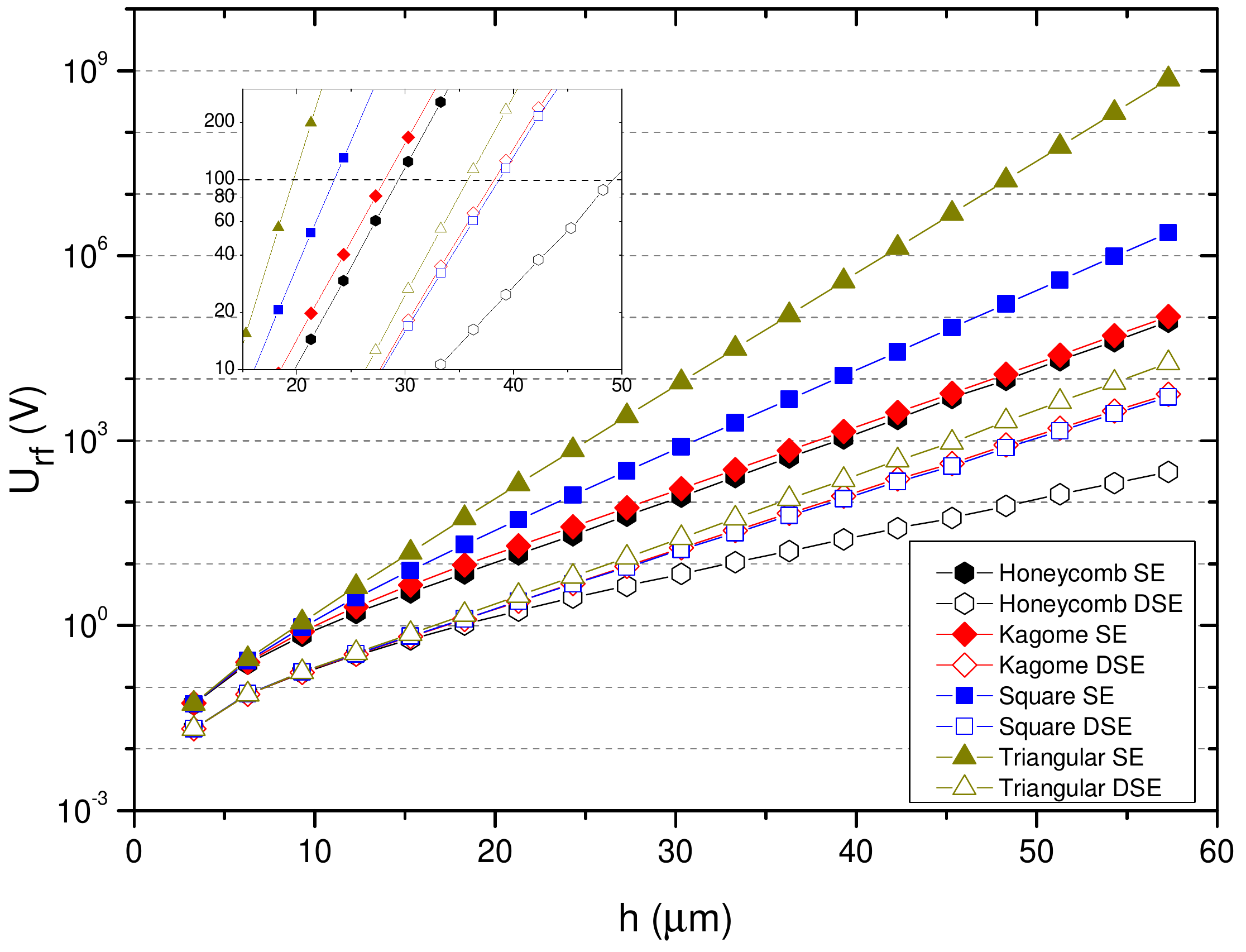}
}
\caption{Radio-frequency voltage amplitude required to reach a mean secular frequency of $5\,\mathrm{MHz}$ in a beryllium trap array with an ion-ion distance of \unit{30}{\micro\meter} and a radio frequency of \unit{50}{\mega\hertz}. The voltage was calculated for the optimal geometries found by maximizing the dimensionless curvature of four different trap lattices at variable height. The values of the voltage for surface-electrode configurations (filled symbols) are significantly higher than in the bi-layer case (open symbols). This is an important advantage of bi-layer traps. The inset is a zoom of the same data into the region of realistic voltages.}
\label{fig:Urf}
\end{figure}

\begin{figure}
\centering
\resizebox{0.8\textwidth}{!}{
  \includegraphics{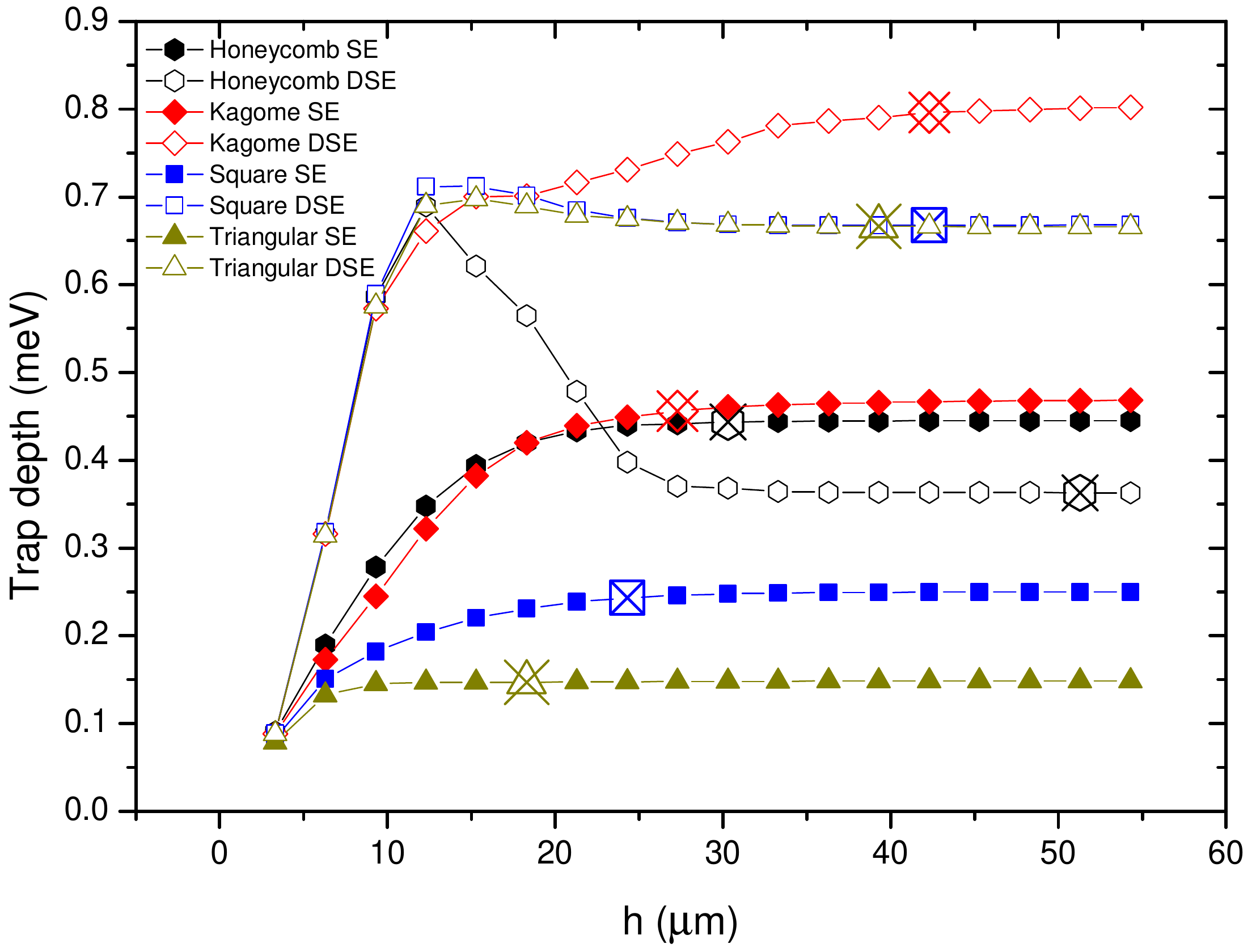}
}
\caption{Trap depth in meV corresponding to the radio-frequency voltages in figure \ref{fig:Urf}. Points to the left of the crossed symbols require voltages below \unit{100}{V}.}
\label{fig:H}
\end{figure}

In order to further quantify the potential impact of bi-layer traps, we choose a \Beplus ion with a mean secular frequency $\tilde{\omega}=2\pi\cdot\unit{5}{\mega\hertz}$ and we fix the ion-ion distance to $d=\unit{30}{\micro\meter}$ and the trap drive frequency $\omega_\text{rf}=2\pi\cdot\unit{50}{\mega\hertz}$, giving an exchange rate $\Omega_\text{ex}\approx2\pi\cdot\unit{3.6}{kHz}$ according to eq. (\ref{eq:Jex}). Figure \ref{fig:Urf} shows the trap drive amplitude $U_\text{rf}$ required to achieve such a configuration for the different lattices in the surface-electrode and bi-layer cases as a function of $h$. For the honeycomb lattice, we see that with \unit{100}{\volt} (dashed line in the inset) we get close to a factor of two increase in the trap height $h$ with respect to the surface-electrode trap. This implies a reduction of more than an order of magnitude in the heating rate of the ion.

Figure \ref{fig:H} shows the trap depth corresponding to the radio-frequency voltages calculated for the plots in figure \ref{fig:Urf}. Here, adding a cover pattern improves the trap depth by factors ranging from 2 to 5, except for the case of the honeycomb lattice, which peaks at a certain value of $h/d$ and even goes below the curve for the single-layer case. The reason for this is that the honeycomb lattice is very efficient at creating curvature in the sense that it requires a comparatively smaller drive amplitude to produce the same secular frequency as other lattices. Shallower traps result from the lower trapping voltages required to achieve a given curvature for this lattice. As a reference, we have marked with large crossed symbols the points for which the radio-frequency voltage is closest to \unit{100}{V} (traps further away from the electrodes require larger voltages in this plot.)

\section{Fabrication considerations}
\label{sec:fab}
\begin{figure}
\centering
\resizebox{0.8\textwidth}{!}{
  \includegraphics{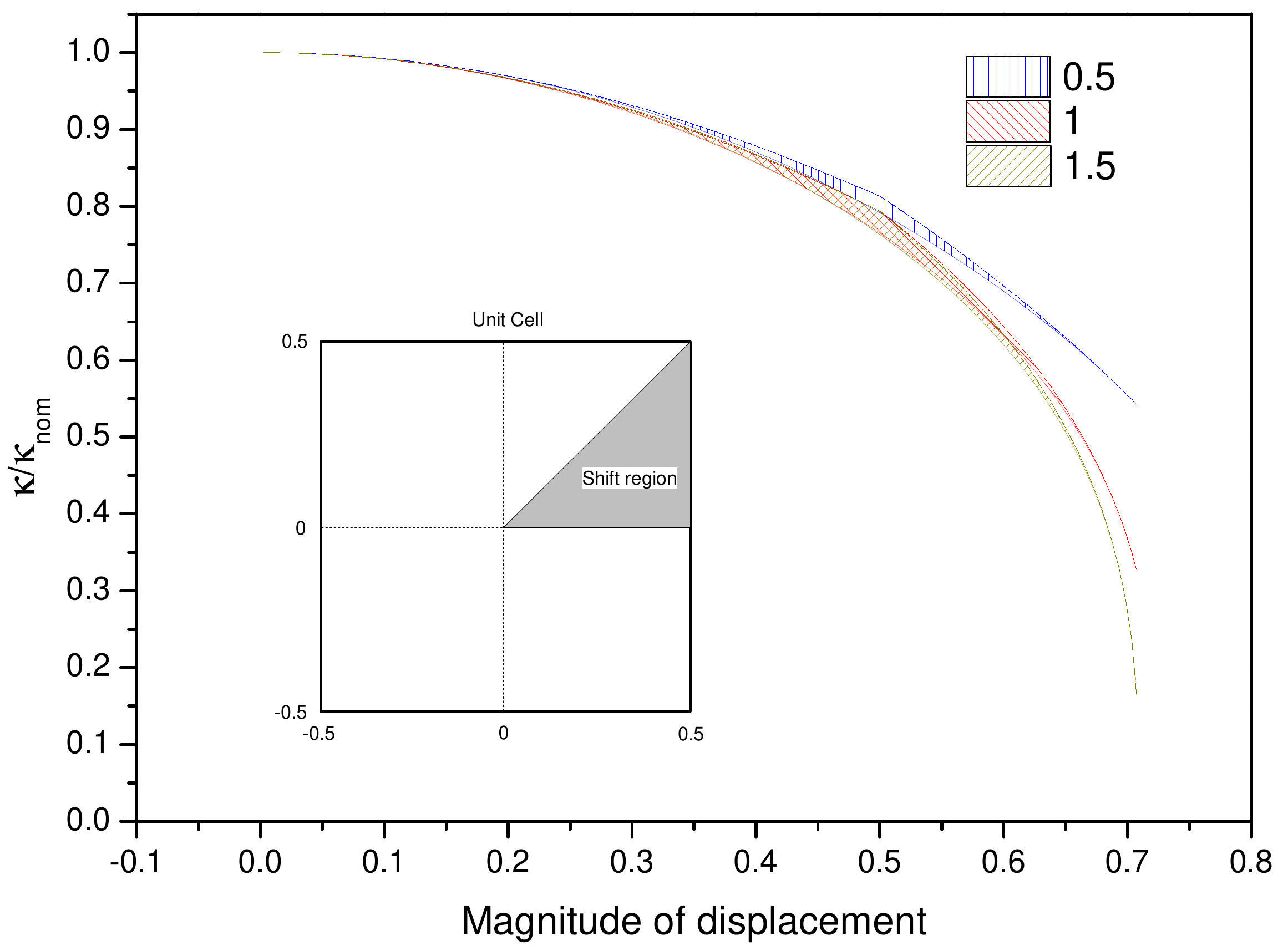}
}
\caption{Dimensionless curvature $\kappa$ of a misaligned square-lattice bi-layer trap as a function of the magnitude of the misalignment in units of the lattice constant. $\kappa$ is given in units of the nominal curvature $\kappa_\text{nom}$ resulting the aligned configuration. Each region corresponds to a different value of $h/d$ (see legend). Inset: square-lattice unit cell and the triangle (half a quadrant) onto which any displacement can be projected.}
\label{fig:kappa_mis}
\end{figure}

\begin{figure}
\centering
\resizebox{0.8\textwidth}{!}{
  \includegraphics{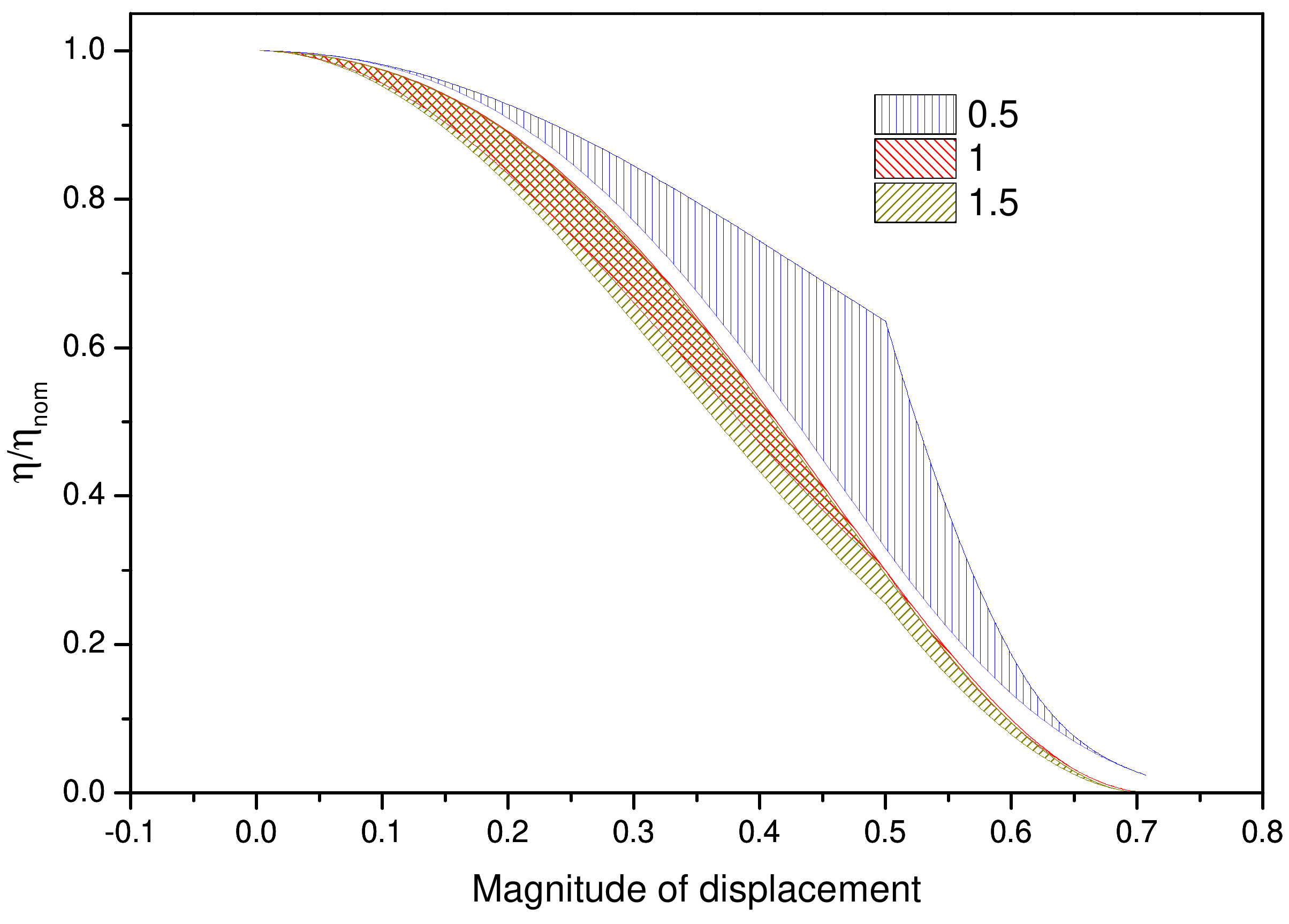}
}
\caption{Dimensionless trap depth $\eta$ of a misaligned square-lattice bi-layer trap as a function of the magnitude of the misalignment in units of the lattice constant. $\eta$ is given in units of the nominal curvature $\eta_\text{nom}$ resulting the aligned configuration. Each shaded curve corresponds to a different value of $h/d$ (see legend).}
\label{fig:etha_mis}
\end{figure}

There are a number of difficulties associated with realizing two-plane structures such as those described above as compared to the single planes which have already been fabricated \cite{11Moehring}. One which has already been found to be challenging in multi-layer electrode traps is the relative alignment of the electrode planes.

In order to gain some insight on the quantitative effect of misalignments, we displaced the upper plane of a square-lattice bi-layer trap and studied the resulting trap lattices. Figures \ref{fig:kappa_mis} and \ref{fig:etha_mis} show the relative reduction of the dimensionless curvature and trap depth respectively as a function of the magnitude of the shifts. Due to the symmetries present in the square lattice, any displacement can be projected onto a displacement within the triangle of half a quadrant of the unit cell (see inset of figure \ref{fig:kappa_mis}). The effect of the shift depends on its direction as well as its magnitude. The boundaries of the shaded area are determined by the directions which produce minimal and maximal changes on the dimensionless quantities. For the case of the square lattice these are along the edges of the triangle. This causes the kinks present in all plots at displacements of 0.5 cell spacings. Note that the calculations account for misalignments of any magnitude in any direction, but not for rotations of one plane with respect to the other. These would lead to position-dependent shifts, meaning that different traps will be affected differently. For tolerances below 10\% of a unit cell, the trap frequencies and depths changes remain very close to the nominal ones. Alignment precision well below \unit{1}{\micro\meter} should be within current experimental reach \cite{11Lee}.

Further foreseeable challenges would be loading such shallow traps (in our example, well below \unit{1}{meV}) or guaranteeing optical access to the array. Regarding the former, a trap depth of \unit{1}{meV} corresponds to less than \unit{12}{K}. Despite the improvement with respect to surface-electrode traps, this imposes difficulties for loading schemes based on ionizing neutral atoms emitted from a hot oven. Instead, other options could be considered, such as loading pre-cooled ions from a magneto-optical trap \cite{12Sage}.

Depending on the ion, optical access could be achieved by building one or both electrode planes with Indium-Tin-Oxide \cite{08Debatin,13Eltony}, which is transparent above \unit{400}{nm}. Another option would be to use integrated photonics \cite{13Eltony,10vanDevender,11Kim,11Brady}.

\section{Discussion}
\label{sec:disc}

Despite the technical challenges discussed above, any reduction on the inter-ion distance is desirable due to the cubic dependence of the ion-ion coupling strength. The smallest ion traps used so far are surface-electrode traps, which can be fabricated with conventional photo-lithographic techniques. The fabrication of bi-layer traps should be similar to that of state-of-the-art surface-electrode traps. Depending on the specific experimental setup, adding the second layer of electrodes (i.e. making the trap 3-dimensional) can yield a gain of over an order of magnitude in coupling strength between neighboring ions when compared to surface-electrode architectures.

\begin{figure}
\centering
\resizebox{1\textwidth}{!}{
  \includegraphics{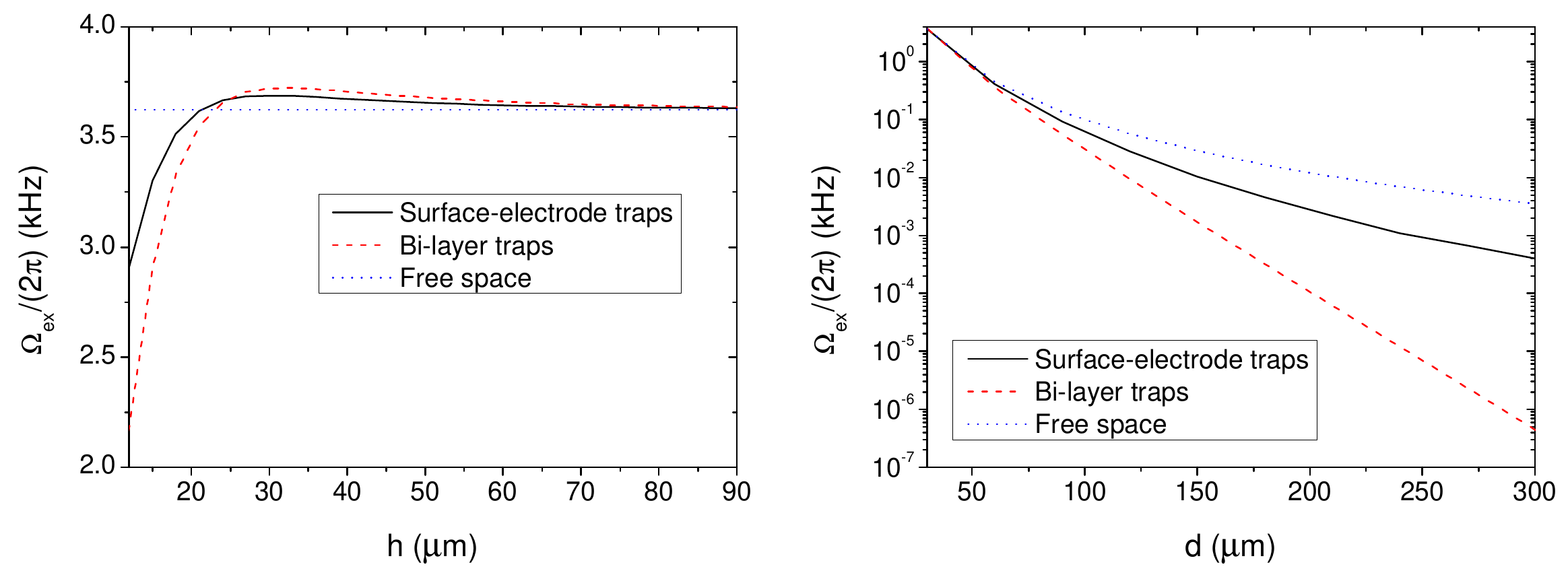}
}
\caption{Energy exchange rate between neighboring \Beplus ions oscillating at \unit{5}{MHz}. Left: Ion separation fixed at \unit{30}{\micro m}. Right: Ion-electrode distance fixed at \unit{30}{\micro m}.}
\label{fig:dipdip}
\end{figure}

One relevant consideration brought up in reference \cite{11Schmied} is that any conducting structure close enough to the ions will affect the interaction between them until the dipole-dipole approximation (eq. (\ref{eq:Jex})) breaks down. In order to see how the coupling strength behaves as a function of $h/d$ for the surface-electrode (SE) and the bi-layer (BL) configurations, we substitute their respective Green's functions,
\be
G_\text{SE}(h,d)\approx\frac{1}{d}-\frac{1}{\sqrt{d^2+4h^2}},\\
G_\text{BL}(h,d)=\displaystyle\sum_{\mu\rightarrow -\infty}^{\infty} G_\text{SE}(h+2\mu H,d),
\ee
in eq. (\ref{eq:Jexgen}). In figure \ref{fig:dipdip} (left) we see they both asymptotically approach the free-space value for large $h/d$, while the coupling is exponentially shielded for ions which are much closer to the electrodes than to their neighbors. For realistic experimental values where $h\approx d$, we see that the ion-ion coupling is enhanced by the conducting planes of electrodes, and more so in the bi-layer case. On the right plot we see how the coupling depends on the inter-ion distance for a fixed value of $h$.

We have also studied the harmonicity of the traps resulting from different periodic electrode structures. Anharmonic traps lead to position-dependent potential curvatures. This can become a problem in the presence of stray electric fields since the interactions become non-resonant. Here we find a further advantage of bi-layer traps when compared to surface-electrode configurations (with or without a grounded cover plane). This arises from the mirror symmetry present in the bi-layer configuration, which makes the trapping potential symmetric along the $z$ axis and therefore removes the strong cubic anharmonic term inherent to surface-electrode traps. We have also analyzed anharmonicities on the lattice plane and observed that the quartic term is considerably weaker in the case of bi-layer traps than for single-plane configurations.

The final issue which we consider is the appearance of unwanted additional trapping sites due to the formation of pseudo-potential minima between the desired traps. These spurious traps will be shallower for small values of $h/d$, in which case they will be difficult to load. The curvature of these sites is also smaller than that of the main traps, so they could be emptied by applying a resonant excitation at one of the motional frequencies. However, these secondary traps can become very similar to the main traps for large values of $h/d$. This depends greatly on the lattice and experimental parameters chosen and should be studied carefully for the particular implementation planned.

\ack

We thank Roman Schmied for many valuable discussions. This work was supported by the Swiss NSF under grant no.~200021\_134776 and the COST programme under grant no.~COST-C12.0118.

\bibliography{myrefs}

\end{document}